\documentclass[sigconf,nonacm]{acmart}

\AtBeginDocument{%
  }

\setcopyright{rightsretained}
\copyrightyear{2025}
\acmYear{2025}
\acmConference[Nyrkiö whitepaper 2025]{Nyrkio whitepaper}{blog.nyrkio.com}
\makeatletter
\renewcommand{\@copyrightowner}{Something special}
\renewcommand{\@acmDOI}{Something special}
\makeatother

\usepackage{graphicx}

\begin{document}

\title{8 Years of Optimizing Apache Otava: How disconnected open source developers took an algorithm from $n^3$ to constant time}

\author{Henrik Ingo}
\email{henrik@nyrki\"o.com}
\orcid{0000-0003-1571-5108} 
\affiliation{%
  \institution{Nyrkiö Oy}
  \city{Järvenpää}
  \country{Finland}
}

%\author{Matt Fleming}
%\email{matt@nyrki\"o.com}
%\affiliation{%
%  \institution{Nyrkiö Ltd}
%  \city{Halifax}
%  \country{United Kingdom}
%}

\renewcommand{\shortauthors}{Nyrkiö whitepaper}

\begin{abstract}
As the project now known as Apache Otava (incubating) makes it first release, we look back over the past 8 years that the codebase was developed by a rather uncoordinated, loosely connected group of performance engineers at MongoDB, Datastax, Confluent, Nyrkiö and others.

Ever since the first publication \cite{MONGOCPD}, developers of the code base now known as Apache Otava (incubating), have continuosly improved its performance. Even when a contributor's primary motivation was to add functionality, it seems like they couldn't help themselves but to also make some performance optimizations while at it.

When developing the Nyrkiö web service to provide change detection for performance testing, we have observed that Otava had become fast enough that it was almost feasible to compute change points synchronously, as the user is browsing test results in a web browser. Inspired by this, we have developed and contributed a new optimization for the common case where new data points are appended to the end of the series. This is now the 7th generation of performance optimizations in Otava. These improvements have been done over the past 8 years of development, by disconnected individuals at different employees. Taken together, the historical optimizations and those published in this paper, represent a 18 000 to 300 000 speedup over the original "by the book" implementation of \cite{EDIV}. In the language of computational complexity, an evolution from $O(n^3)$ to $O(1)$ (constant time). The ability to compute and recompute change points in real-time unlocks new opportunities in the user experience.
\end{abstract}
\keywords{change point detection, performance, benchmarking, continuous
integration}
\begin{teaserfigure}
  \includegraphics[width=\textwidth]{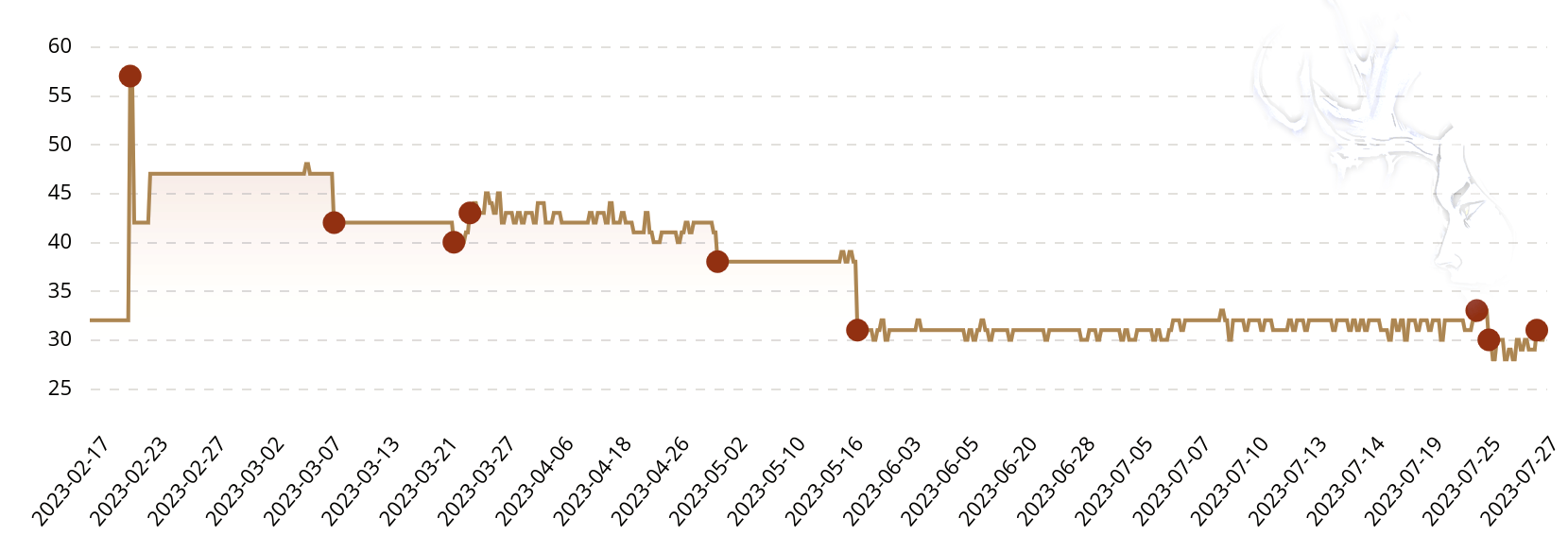}
  \caption{Change points in a demo data set from TigerBeetle}
  \Description{The forest spiŕit Nyyrikki tirelessly hunting for performance changes.}
  \label{fig:teaser}
\end{teaserfigure}

\maketitle

\section{Introduction}

\subsection{Historical review}

Using Change Point Detection, more precisely the E-Divisive Means algorithm\cite{EDIV}, to analyze benchmarking results, was first introduced by the performance team at MongoDB in 2017. In 2020 This work was both published as open source\cite{MONGOSIG}, and reported on in a conference paper.\cite{MONGOCPD} 

Soon after both Netflix\cite{NETFLIXBLOG} and Datastax\cite{Hunter} reported that they had implemented a similar system, in their Continuous Integration (CI) pipelines,based on the python library open sourced by MongoDB. Again with positive results, preventing performance regressions in production releases. Datastax published their additions and improvements as the open source Hunter project.\cite{HUNTERREPO} Hunter expanded the original MongoDB Python library\cite{MONGOSIG} into a fully functional command line tool, that integrates with the necessary tools in a performance engineer's workflow, such as Grafana annotations and Slack notifications. It also further improved both the accuracy and performance over the MongoDB approach, mostly thanks to changing the statistical significance test to use Student's T test.

The availability of a full featured command line tool further lowered  the bar for adopting change point detection for the analysis of a \emph{Continuous Performance Engineering} pipeline. In 2024 Confluent reported on their adoption of the tool, then still known as Hunter, to process nightly results from their Kafka performance tests.\cite{AUTOSPEED}

% The presentation includes an interesting summary of the investment needed for setting up and using Hunter based Change Point Detection: 3 engineering months for setup, and after that, a mere 389 USD in monthly infrastructure costs. The long setup investment is because in practice all of these in-house change point detection systems are not just a command line tool, rather the benchmark results are stored in a timeseries database, presented for inspection with some graphing solution (e.g. Grafana dashboards, or some in house solution). Typically there's a need to integrate with several other tools used to manage and fix regressions: Grafana, Jira, Github, Slack and email.

The author likewise continues to build upon and use this code base in their professional work. Even if during the past 8 years change point detection had become a necessary component in the toolkit of the performance engineering teams, at least the ones this author has worked in, it was nevertheless a side project and a means to an end. This has changed in 2024 when the author has founded a startup to productize and commercialize change point detection. What was once a side project to solve a specific problem, has now become a full time occupation, and a mission, to bring change point detection, and the whole concept of \emph{Continuous Performance Engineering} to the mainstream.

While building the Nyrkiö web service platform around Apache Otava (incubating), we noticed that the performance improvements introduced in \cite{Hunter} had brought the change point detection code close to a speed that makes it a realistic goal to compute change points on demand, as the user is reviewing the performance test results in a web browser. Inspired by this we have introduced some performance optimizations of our own to truly minimize the computational cost.

As the use of Change Point Detection is spreading across the industry, Datastax in 2024 donated the code base to the Apache Software Foundation. Going forward the project will be known as Apache Otava (incubating). Being part of the Apache ecosystem has again brought more awareness and validation to using e-divisive means for automated detection of change points in a \emph{Continuous Performance Engineering} pipeline.

The process of incorporating the loose network of forks into an official Apache open source project, has brought together engineers from MongoDB, Datastax and other comopanies, that over the past 8 years contributed rather loosely, in an on-off fashion to the project now known as Apache Otava (incubating). To celebrate 8 years of open source contributions from a dozen individuals working on solving the same engineering challenge at different employers, the focus on this paper will be specifically on performance improvements made to the algorithm itself. When combining both historical improvements and new ones first presented in this paper, we will recount the rather remarkable journey of how the e-divisive algorithm implemented in Apache Otava was optimized from a problematic $O(n^3)$ to constant time $O(1)$.

/subsection{Structure of the remaining article}

For the benefit of the reader, we will in Section 2 first briefly review how the e-divisive algorithm works, and a typical workflow of using Apache Otava (incubating) to automate detection of regressions in a \emph{Continuous Performance Engineering} pipeline. These have been described in more detail in \cite{EDIV}, \cite{MONGOCPD}, \cite{DSI} and \cite{Hunter}.

In Section 3 we review the performance optimizations that were implemented by the MongoDB and Datastax teams in \cite{MONGOCPD} and \cite{Hunter}, respectively. Taken together, this prior work already improved the performance of the implementation in Apache Otava by 8500 x compared to the initial naive implementation. 

In Section 4 we augment the analysis of the computational complexity of the previously reported results. In the case of the MongoDB paper\cite{MONGOCPD}, such an analysis is provided, but it missed the fact that the most significant contributor to the computational cost is the Monte Carlo significance test that happens at the end of the algorithm. We add this to the $O()$ notation. In the case of the Datastax paper\cite{Hunter} the authors were focused on qualitative improvements to the accuracy of the algorithm and did not provide an $O( )$ notation analysis at all. We contribute one here.

Section 5 evaluates the significance of these improvements to the user experience of our web service, when change points can be recomputed efficiently, on demand. It was the observation that Otava had become fast enough to achieve this, that inspired us to pursue a further optimization, which is presented in Section 6.

\section{E-divisive Change Point Detection for Dummies}

\begin{figure}
    \centering
    \includegraphics[width=0.5\linewidth]{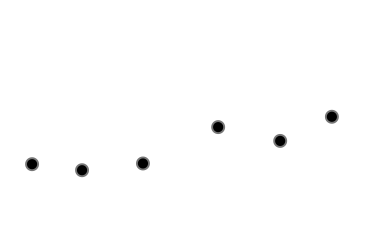}
    \caption{A short timeseries used for illustration purposes}
    \label{fig:dummies1}
\end{figure}

To briefly explain the high level idea of the e-divisive algorithm \cite{EDIV} let's consider the short series of 6 benchmarking results over some time period, as illustrated in Figure \ref{fig:dummies1} What you would want to know, is whether the slight variations from one point to the next are just random noise, or represent an actual preformance regression in the code being tested.

While this is a somewhat easy task for the human eye to spot - in this figure it appears like there might be a real change between the third and fourth points - it is a surprisingly hard task to answer in an automated fashion. Yet we need automation to achieve scale, because asking a human performance expert to browse through hundreds of graphs every morning is not realistic. 

Some naive attempts at creating automated alerts, will make the situation worse: Setting the alert at some large treshold, like  10\% will not help, because there will be larger spikes also in the random noise. Thus such an alert will miss small real regressions, yet still produce false positives at an annoying rate. Further, each benchmark and metric will have their own unique pattern of random noise. For some, the range is 3-4\%, for other tests it could be 15-20\%. Another naive solution that comes to mind is then to parameterize the alert thresholds so that they are configurable per test. The "automated" solution then turns out to require a surprising amount of manual handholding, if the performance engineer is to figure out an optimal threshold for each test separately. In conclusion, a more sophisticated statistical method is needed.

Immediately at the start, the e-divisive algorithm takes a different approach than most other algorithms we reviewed or thought of. Commonly the problem of validating results from a \emph{Continuous Benchmarking} pipeline is seen as one where one or a few newest results are compared against a history of results. E-divisive on the other hand starts with the entire history, and will look for statistically significant changes \emph{anywhere in the history provided}. It turns out this approach is the reason E-Divisive achieves such good accuracy: It maximally utilizes the information available. Where other approaches compare a single point to its history, e-divisive will consider each point and take into account both its history and future!

\begin{figure}
    \centering
    \includegraphics[width=0.5\linewidth]{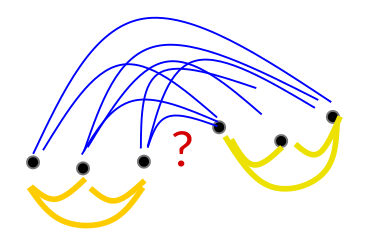}
    \caption{Pairwise differences}
    \label{fig:dummies2}
\end{figure}

\begin{figure*}
  \includegraphics[width=\textwidth,height=4cm]{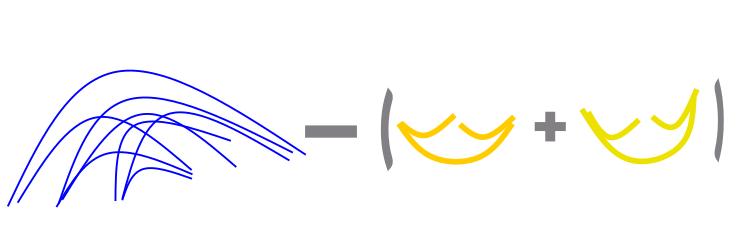}
  \caption{The E-divisive algorithm, illustrated. (See \cite{EDIV} for actual math.}
  \label{fig:dummies3}
\end{figure*}

The algorithm starts by computing the \emph{pairwise differences} of all points in the series, as illustrated in figure \ref{fig:dummies2}. It then steps through each point, or maybe rather each gap between the points, to compute a value denoted as $\hat{q}$. This is computed by summing together all the differences across the "gap" in question, and then subtracting all the pairwise differences where both points of the pair are on the same side of the "gap". (Figure \ref{fig:dummies3}) Intuitively, the idea is that if the difference between two consecutive points is "real", that is, \emph{persistent} and \emph{statistically significant}, then the sum of differences that includes the gap in question will be larger than the sum of random variations on each side. On the other hand if the gap we observe is just a symptom of random noise, then both sides of the subtraction will be roughly equal, yielding a $\hat{q}$ value close to zero.

\begin{figure}
    \centering
    \includegraphics[width=0.5\linewidth]{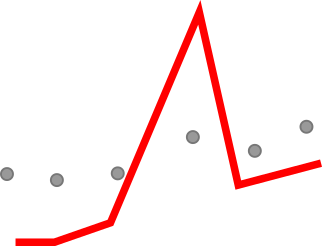}
    \caption{$\hat{q}$ computed for each gap between points in the series}
    \label{fig:dummies4}
\end{figure}

Computing the above $\hat{q}$ for each "gap" in the series, will yield a continuous graph (figure \ref{fig:dummies4}) where $\hat{q}$ is larger, the bigger and more persistent a change is. Therefore, to detect a change point, we simply search for the maximum of the graph in figure \ref{fig:dummies4}.

The remaining step is to design whether this "largest" change is truly a real change (a real performance regression, in our case) or whether it just happens to be the largest exhibit of random noise in the series. For this, we employ a \emph{statistical significance test}. In the original version of the algorithm, the significance is computed with a Monte Carlo simulation: The points in the series are shuffled, then all the entire algorithm is run again. The idea here is that if the change at the gap in question is just due to random variation, then in a random shuffle there will also exist gaps where the $\hat{q}$ value is just as large or larger. While this Monte Carlo approach is both easy to comprehend and fun to implement, as we will see, it is also both slow and inaccurate at the same time.

Finally, the e-divisive algorithm is \emph{iterative}. So if the chosen $max(\hat{q})$ is determined to be a real (statistically significant) change point, then we take the left and right halves of the time series separately and start again with a new round of e-divisive for both halves. This continues until the statistical significance test fails, at which point we determine that no new change points could be found in the remaining parts of the series.

\section{Performance optimizations implemented previously}

Both \cite{MONGOCPD} and \cite{Hunter} made significant performance improvements to the E-Divisive algorithm and its Python implementation. To recap, the significant improvements of each have been summarized in Table \ref{tab:perfimprovements}.

\begin{table}
    \centering
    \begin{tabular}{|c|c|c|}
      \hline
        Version                  & Complexity & Relative speed \\
      \hline
        Naive                    & $O(\kappa T^3 m)$ & 6103 \\
        Shift row/col difference & $O(\kappa T^2 m)$ & 113 \\
        Numpy                    & $O(\kappa T^2 m)$ & 13 \\
        Native (C)               & $O(\kappa T^2 m)$ & 1 \\
      \hline
        Hunter improvements & & \\
      \hline 
        Student's T test             & $O(\kappa T^2$)  &  0.838 \\
        Student + W                  & $O(TW)$  &  0.03 - 0.32 \\
      \hline
        New improvements & & \\
      \hline
        Incremental                  & $O(W^2)$ &  0.02 - 0.25 \\
      \hline
    \end{tabular}
    \caption{Performance improvements previously achieved in the E-Divisive Python implementation}
    \label{tab:perfimprovements}
\end{table}

Note that while \cite{Hunter} introduced the Student's T-test and Weak Change Points modifications as qualitative improvements, they both actually brought significant performance improvements too. While this was benchmarked in the paper, a discussion on computational complexity was omitted. We will explain it here.

A significant contributor to the computational complexity of the original E-Divisive implementation comes from the Monte Carlo significance test. It is expensive, because it requires to redo the entire algorithm from the start with $m$ permutations of the series being tested. In fact, because it is the dominating component of execution time, all practical implementations of E-Divisive we have seen, have used 100 permutations only. However, this is a low number. Since the p-value is often expressed as a percentage, it means that the outcome of a single shuffle may determine whether a candidate change point was considered to be statistically significant or not! Further, it also means that 0.01 is the smallest non-zero p-value we can even use, since the granularity of the significance test doesn't allow smaller decimals than that. Hence, in practice a more robust value for $m$, when using the original implementation of E-Divisive\cite{MONGOCPD}, would be at least $m=1000$ if not $m=10000$. However, this would explode computation times from seconds to minutes, and even hours in the latter case, which would hurt the user experience as processing the results would take longer than the benchmark itself.

In conclusion, replacing the original Monte Carlo simulation with a common Student's T-test, was therefore a significant performance improvement in Hunter\cite{Hunter}. In practice we have found it to be even one or two orders of magnitude, not just 18\% that is the improvement reported in the Datastax paper. The approaches scale differently with the number of change points $\kappa$ found, and in practice we often find less than the 20 found in the benchmarks in \cite{Hunter}.

\section{Computational Complexity Analysis of the Historical Results}

To emphasize the dominating impact of the number of permutations to the computational complexity of E-Divisive, we have in Table 1 added $m$ to the first four versions. We note that the empirical results are correct and unchanged in our table.

The computational complexity of Student's T-test is merely that of computing an average and standard deviation, both of which are $O(T)$. Since this is then no longer the dominating factor, we have omitted it from the following rows in the table.

The last modification in \cite{Hunter} that has a material performance impact was motivated by wanting to make the algorithm more sensitive to short lived changes. If a regression is introduced and then quickly fixed, Hunter will find two change points where the original algorithm would typically ignore at least one, if not both of them. The modification that achieves this is to break the timeseries into multiple shorter windows. This causes the algorithm to be more likely to find such pairs of change points. \footnote{Intuitively, a deviation lasting two measurements--maybe 2 days--in a series of 50 measurements (days), is 4\% of the series, while the same two measurements in a series of 300 measurements start to look like a passing anomaly. This is exactly how E-Divisive will judge them too.}

While that modification superficially appears to make the computation more complex, it actually reduces the dominating part. We no longer need to compute the large difference matrix between all data points $T$, which is the $T^2$ factor. Instead we compute separately a small window of size $W$ (typically $W=50$) and we do that $2 T/W$ times. (Or $\kappa$, in the odd case that it is greater.) We therefore get $O(W^2 T/W) = O(T W)$. Again, in our own experience we find that the improvement is often larger than reported in \cite{Hunter}. This would follow theoretically if $T$ is large, but also when the number of change points found ($\kappa$) is large.

Hence, the optimizations that are almost a "by-product" in Hunter have brought the algorithm down to linear time (as $W$ is constant). \footnote{A fact missed by the authors of \cite{Hunter}, but worth celebrating!}

\section{Practical benefit of using a fast Change Point Detection algorithm}

For Nyrkiö we recently built a web service that wraps Apache Otava. Our service offers a HTTP API endpoint to add (or delete) performance test results, a MongoDB database to store them, and a lot of JavaScript to present them in nice graphs. In all previous implementations of E-Divisive for analyzing performance testing results, that we know of or have participated in building, the computation of change points was implemented as a background task. The computation will take seconds, and is expensive enough that redoing it on every page load would be wasteful.

Or so we thought! As we started building the Change Point Detection service that we now work on, it was initially thought of as a temporary solution to not implement a work queue, rather we just executed Otava on every page load! We were surprised to realize this was almost realistic to do! We kept developing and even demoed the site several times with Otava running on demand, no background pre-computing needed. Seeing that this was possible was the inspiration to better analyze the performance improvements in Otava, which we have now done.

Eventually we did start storing the change points rather than computing them again on every page load. One reason is that we may display dozens of timeseries on a single page, at which point the total time to compute all change points becomes seconds or even a minute or two. This could be easily solved by using servers with slightly more CPU, but that seemed wasteful as a long term solution. Also, we offer services in the area of software performance - we cannot afford for our own website to appear slow. A final reason was that we want to show summary statistics computed from all recent change points on a front page and it would simply not be possible to recompute all the change points of the entire site every time someone views the front page. 

The idea of first computing a set of "weak change points", that is, a superset of the correct set of change points, computed by setting $m$ to a large value, can also be used to speed up the user interface. We allow users to tune the key parameters of the algorithm: \emph{p-value} and \emph{minimum magnitude} of the change. Any time a user changes these values in a configuration panel, we must recompute the change points. But if we store the set of weak change points, we don't need to recompute them, rather now we can simply redo the filtering from weak change points to actual change points. (See \emph{merge()} in \cite{HUNTERREPO}.) In this case we don't need to recompute anything at all, rather we can react almost instantaneously to the user sliding the configuration widgets. In fact, displaying or hiding change points based on user input could even be done purely in the frontend.

\section{Incremental Otava}

Having finally implemented the persisting of computed change points, this opens the door for one more performance optimization: The typical case is that  new data points are added to the end of the timeseries. Using the Hunter modifications to E-Divisive, we already split the timeseries to smaller windows of 30 points each. If it is the case that a new point was added to the end, we do not need to re-compute any of the other windows. It is sufficient to recompute 1-2 window length's worth of points from the end of the series. Since at any given point, the algorithm will only consider a window of W points, the newly appended last point cannot have any effects on the points that are earlier than the last W points.

The computational complexity of this common case is now reduced to a constant $O(W^2)$. Benchmark results using a 365 point demo dataset from TigerBeetle performance testing are presented in Table 2. As discussed in Section 3, the different versions of the algorithm scale differently wrt number of change points found. For this reason we ran the benchmark with different $m$-values.

In addition we have added a representation of this optimization to the end of Table 1. Depending on the number of change points found, which in turn depends on the p-value chosen, the incremental Otava version is between 4x and 50x faster than the "Native C" version. We wish to highlight that altogether these performance improvements have now made Otava 18 thousand to 300 thousand times faster than the first "Naive" version that implemented E-Divisive "by the book".

The benchmark was added to and executed with the standard Otava pytest command on a Dell XPS 13 laptop with a i7-6560U CPU @ 2.20GHz, 2 cores x 2 hyper threads. $numactl -C$ was used to pin the execution to a single CPU core and thread.

Since this optimization is only for the incremental case, the benchmark setup was such that Otava was executed twice: First a regular full pass over the 365 point series, then a single point was added to the end of the series, after which we recompute the change points in incremental mode. We then subtract the execution time of the first, single pass, and report only the time it takes to compute the second pass.

\begin{table}
    \centering
    \begin{tabular}{|c|c|c|c|c|}
    \hline
                          & p=0.001 & p=0.01 & p=0.1 & p=0.2 \\
      Change points found &       6 &     9  &    31 &    47 \\
    \hline
      Native (C)          &   253.0 &  345.0 &   387 &  565  \\
      Student + W         &     8.2 &   28.1 &   184 &  181  \\
      Incremental         &     5.5 &   18.0 &   157 &  144  \\
    \hline
      Relative improvement&    &  &   &   \\
      Student + W         &  0.03	&  0.08 &  	0.40 & 0.32  \\
      Incremental         &  0.02	&  0.05	&   0.41 &	0.25    \\
    \hline
    \end{tabular}
    \caption{Benchmark results on a 365 point demo dataset from TigerBeetle performance tests. We report the median value from 100 runs. Values in the middle section are in millisecond and at the bottom relative to the reference implementation "Native (C)".}
    \label{tab:my_results}
\end{table}

\section{Conclusions}

Change Point Detection is continuing to spread in the Continuous Performance Engineering community. The open sourcing of a fully functional command line tool\cite{HUNTERREPO} has further lowered the threshold for more companies to adopt this technique. The author has launched a startup with the aim of bringing change point detection into the mainstream. The code base that implements the e-divisive algorithm was donated by Datastax to the  Apache Software Foundation, and is now known as Apache Otava (incubating project).

The original paper on using E-Divisive to analyze benchmarking results\cite{MONGOCPD} had not included the cost of the rather heavy Monte Carlo simulation at the end of computing a change point. We corrected this by adding the number of permutations $m$ to the $O(...)$ notation.\footnote{Note that even if we assign blame to the third person due to the customary writing style, it is the same third person as is authoring the present article.} The Hunter paper\cite{Hunter} did not include an $O(...)$ analysis of its algorithmic improvements, so we have contributed them here.

We introduce a new, powerful optimization by noting that the common case is to add points to the end of the series, and in this case it is sufficient to recompute only the last 1-2 windows of the series.

Taken together, the prior and present optimizations have made Otava fast enough that we can re-compute change points in real-time, when the user adds a new test result, or changes the parameters of the algorithm, interactively.

\bibliographystyle{ACM-Reference-Format}
\bibliography{8_Years_of_Optimizing_Apache_Otava}


\begin{thebibliography}{8}

%%% the Web address.

\ifx \showCODEN    \undefined \def \showCODEN     #1{\unskip}     \fi
\ifx \showDOI      \undefined \def \showDOI       #1{#1}\fi
\ifx \showISBNx    \undefined \def \showISBNx     #1{\unskip}     \fi
\ifx \showISBNxiii \undefined \def \showISBNxiii  #1{\unskip}     \fi
\ifx \showISSN     \undefined \def \showISSN      #1{\unskip}     \fi
\ifx \showLCCN     \undefined \def \showLCCN      #1{\unskip}     \fi
\ifx \shownote     \undefined \def \shownote      #1{#1}          \fi
\ifx \showarticletitle \undefined \def \showarticletitle #1{#1}   \fi
\ifx \showURL      \undefined \def \showURL       {\relax}        \fi
% invisible to TeX
\providecommand\bibfield[2]{#2}
\providecommand\bibinfo[2]{#2}
\providecommand\natexlab[1]{#1}
\providecommand\showeprint[2][]{arXiv:#2}

\bibitem[Daly et~al\mbox{.}(2020)]%
        {MONGOCPD}
\bibfield{author}{\bibinfo{person}{David Daly}, \bibinfo{person}{William Brown}, \bibinfo{person}{Henrik Ingo}, \bibinfo{person}{Jim O’Leary}, {and} \bibinfo{person}{David Bradford.}} \bibinfo{year}{2020}\natexlab{}.
\newblock \showarticletitle{The Use of Change Point Detection to Identify Software Performance Regressions in a Continuous Integration System}.
\newblock \bibinfo{journal}{\emph{In Proceedings of the 2020 ACM/SPEC International Conference on Performance Engineering(ICPE ’20)}} (\bibinfo{year}{2020}).
\newblock
\urldef\tempurl%
\url{https://doi.org/10.1145/3358960.3375791}
\showDOI{\tempurl}


\bibitem[Fleming et~al\mbox{.}(2023)]%
        {Hunter}
\bibfield{author}{\bibinfo{person}{Matt Fleming}, \bibinfo{person}{Piotr Kolaczkowski}, \bibinfo{person}{Ishita Kumar}, \bibinfo{person}{Shaunak Das}, \bibinfo{person}{Sean McCarthy}, \bibinfo{person}{Pushkala Pattabhiraman}, {and} \bibinfo{person}{Henrik Ingo}.} \bibinfo{year}{2023}\natexlab{}.
\newblock \showarticletitle{Hunter: Using Change Point Detection to Hunt for Performance Regressions}. In \bibinfo{booktitle}{\emph{Proceedings of the 2023 ACM/SPEC International Conference on Performance Engineering}} (Coimbra, Portugal) \emph{(\bibinfo{series}{ICPE '23})}. \bibinfo{publisher}{Association for Computing Machinery}, \bibinfo{address}{New York, NY, USA}, \bibinfo{pages}{199–206}.
\newblock
\showISBNx{9798400700682}
\urldef\tempurl%
\url{https://doi.org/10.1145/3578244.3583719}
\showDOI{\tempurl}


\bibitem[Ingo and Daly(2020)]%
        {DSI}
\bibfield{author}{\bibinfo{person}{Henrik Ingo} {and} \bibinfo{person}{David Daly}.} \bibinfo{year}{2020}\natexlab{}.
\newblock \showarticletitle{Automated System Performance Testing at MongoDB}.
\newblock \bibinfo{journal}{\emph{In Workshop on Testing Database Systems (DBTest’20)}} (\bibinfo{year}{2020}).
\newblock
\urldef\tempurl%
\url{https://doi.org/10.1145/3395032.3395323}
\showDOI{\tempurl}


\bibitem[Kroll(2022)]%
        {NETFLIXBLOG}
\bibfield{author}{\bibinfo{person}{Angus Kroll}.} \bibinfo{year}{2022}\natexlab{}.
\newblock \showarticletitle{Fixing Performance Regressions Before They Happen}.
\newblock \bibinfo{journal}{\emph{...}} (\bibinfo{year}{2022}).
\newblock
\urldef\tempurl%
\url{https://netflixtechblog.com/fixing-performance-regressions-before-they-happen-eab2602b86fe}
\showURL{%
\tempurl}
\newblock
\shownote{blog}.


\bibitem[Matteson and James(2014)]%
        {EDIV}
\bibfield{author}{\bibinfo{person}{David~S. Matteson} {and} \bibinfo{person}{Nicholas~A. James}.} \bibinfo{year}{2014}\natexlab{}.
\newblock \showarticletitle{A Nonparametric Approach for Multiple Change Point Analysis of Multivariate Data}.
\newblock \bibinfo{journal}{\emph{J. Amer. Statist. Assoc.}} \bibinfo{volume}{109}, \bibinfo{number}{505} (\bibinfo{year}{2014}), \bibinfo{pages}{334--345}.
\newblock
\showISSN{01621459}
\urldef\tempurl%
\url{http://www.jstor.org/stable/24247158}
\showURL{%
\tempurl}


\bibitem[MongoDB({[n.\,d.]})]%
        {MONGOSIG}
\bibfield{author}{\bibinfo{person}{MongoDB}.} \bibinfo{year}{[n.\,d.]}\natexlab{}.
\newblock \bibinfo{title}{Signal Processing Algorithms}.
\newblock
\newblock
\urldef\tempurl%
\url{https://github.com/mongodb/signal-processing-algorithms}
\showURL{%
\tempurl}
\newblock
\shownote{Accessed: 2021-10-13}.


\bibitem[{Piotr Kołaczkowski}({[n.\,d.]})]%
        {HUNTERREPO}
\bibfield{author}{\bibinfo{person}{{Piotr Kołaczkowski}}.} \bibinfo{year}{[n.\,d.]}\natexlab{}.
\newblock \bibinfo{title}{Hunter – Hunts Performance Regressions}.
\newblock
\newblock
\urldef\tempurl%
\url{https://github.com/datastax-labs/hunter}
\showURL{%
\tempurl}
\newblock
\shownote{Accessed: 2022-12-12}.


\bibitem[Roesler(2024)]%
        {AUTOSPEED}
\bibfield{author}{\bibinfo{person}{Alex Sorokoumov and John Roesler}.} \bibinfo{year}{2024}\natexlab{}.
\newblock \showarticletitle{Automating Speed: A proven approach to preventing performance regressions in Kafka Streams}. In \bibinfo{booktitle}{\emph{Kafka Summit London}}.
\newblock
\urldef\tempurl%
\url{https://www.confluent.io/events/kafka-summit-london-2024/automating-speed-a-proven-approach-to-preventing-performance-regressions-in/}
\showURL{%
\tempurl}


\end{thebibliography}

\end{document}